\documentclass[trackchanges,twocolumn]{aastex701}
\setcitestyle{authoryear,open={(},close={)}}

\UseRawInputEncoding

\usepackage{comment}
\usepackage{xcolor}
\usepackage{amsmath}
\usepackage{appendix}
\usepackage{footmisc}
\usepackage{subfigure}
\usepackage[utf8]{inputenc}

\defcitealias{zhang_hot_2024}{Z24a}
\defcitealias{zhang_hot_2024-1}{Z24b}
\defcitealias{hopkins_cosmic_2025}{H25}
\defcitealias{das_thermal_2025}{D25}

\shorttitle{The Case for CRs: A CR-Supported L$^{\ast}$ Halo Paradigm}
\shortauthors{Ponnada et al.}
\begin{document}

\title{Strong Evidence for Cosmic Ray-Supported $\sim$L$^{\ast}$ Galaxy Halos via X-ray \& tSZ Constraints}

\author[0000-0002-7484-2695]{Sam B. Ponnada}\email{sponnada@caltech.edu}
\affiliation{Cahill Center for Astronomy and Astrophysics, California Institute of Technology, Pasadena, CA 91125, USA}
\correspondingauthor{Sam B. Ponnada}

\author[0000-0003-3729-1684]{Philip F. Hopkins}\email{phopkins@caltech.edu}
\affiliation{Cahill Center for Astronomy and Astrophysics, California Institute of Technology, Pasadena, CA 91125, USA}

\author[0009-0001-6243-0861]{Yue Samuel Lu}\email{yul232@ucsd.edu}
\affiliation{Department of Astronomy and Astrophysics, University of California, San Diego, La Jolla, CA 92093, USA}

\author[0000-0002-1616-5649]{Emily M. Silich}\email{esilich@caltech.edu}
\affiliation{Cahill Center for Astronomy and Astrophysics, California Institute of Technology, Pasadena, CA 91125, USA}

\author[0000-0003-1257-5007]{Iryna S. Butsky}\email{iryna.butsky@gmail.com}\thanks{NASA Hubble Fellow}
\affiliation{Kavli Institute for Particle Astrophysics and Cosmology, and Department of Physics, 
 Stanford University, 
 Stanford, CA 94305, USA}

\author{Du\v{s}an Kere\v{s}}\email{dkeres@physics.ucsd.edu}
\affiliation{Department of Astronomy and Astrophysics, University of California, San Diego, La Jolla, CA 92093, USA}
\affiliation{Department of Physics, 
University of California 
San Diego, La Jolla, CA 92093, USA 
}

\begin{abstract}

Many state-of-the-art galaxy simulations featuring traditional feedback modes have significant challenges producing enough extended soft X-ray ($\sim 0.5-2$ keV) emission at R $\sim 0.5-1$ R$_{\rm vir}$ observed around galaxies with stellar masses M$_{\rm \ast} \lesssim 10^{11} \rm M_\odot$,  without violating galaxy mass function constraints. Moreover, thermal Sunyaev-Zel'dovich (tSZ) measurements probing the thermal pressure of similar galaxies indicate it is \textit{orders-of-magnitude lower} than predictions from simple halo hydrodynamics and many hydrodynamical simulations. We demonstrate that these constraints can be met congruously with a large non-thermal pressure contribution in the form of cosmic rays (CRs) from SNe and/or AGN, which lowers the tSZ signal while CR leptons produce plentiful soft X-rays via inverse Compton scattering of the CMB. The combination of these two observations is far more constraining on the pressure budget of galactic halos than either alone -- if these novel tSZ and X-ray observations are borne out by future studies, then taken together they reveal \textit{the strongest evidence for CR support in halos to date}. Conversely, it is very difficult to produce the extended X-rays via traditional thermal emission without increasing the overall thermal pressure and thus tSZ signal in tandem, making these tensions even worse. Finally, tSZ \& X-rays together unlock a novel observational method to constrain halo CR pressure relative to thermal pressure, with implications for CR transport parameters and AGN feedback energetics across various galaxy mass scales. Taking the currently observed constraints at M$_{\rm halo} \sim 10^{\rm 12} \rm M_\odot$ imply the halo CR pressure must \textit{at least} be equal to the gas thermal  pressure.

\end{abstract}



\section{Introduction}

\setcounter{footnote}{0}

Interest has been abuzz on the role of cosmic rays (CRs) in galaxy formation in the past decade, which in part owes to puzzling observations probing the circumgalactic medium (CGM). Optical/ultra-violet spectroscopic surveys targeting halos around low redshift $\sim$L$^{\ast}$ galaxies\footnote{Here, loosely defined as galaxies residing in dark matter halos of $M_{\rm DM\, halo} \sim 10^{12} - 10^{13} M_\odot$ \citep{girelli_stellar--halo_2020}.} ($z \lesssim 0.35$), both actively star-forming and quenched, have found that ``cool" gas (T $\sim$ 10$^{4}$ K) is ubiquitous in the CGM \citep{Tumlinson2017}. Moreover, the densities of this cool gas were inferred to be orders-of-magnitude lower than naive expectations for cool clouds in thermal pressure equilibrium with the hot, virialized CGM phase (T $\gtrsim$ 10$^{5.5}$ K for $\sim$L$^{\ast}$ galaxies; \citealt{werk_cos-halos_2014}), implying the existence of substantial non-thermal pressure support.

Our understanding of the multi-phase CGM has progressed via recent stacked observations of $\sim$L$^{\ast}$ galaxies reporting extended, diffuse CGM X-ray emission (\citealt{zhang_hot_2024,zhang_hot_2024-1}; hereafter \citetalias{zhang_hot_2024} \& \citetalias{zhang_hot_2024-1}). Notably, these extended profiles are typically \textit{under-predicted} by state-of-the-art, cosmological simulations of galaxy formation, employing a variety of stellar and AGN feedback schemes \citep{truong_x-ray_2023,silich_x-ray_2025,zhang_tracing_2025}, and also in apparent disagreement with upper-limits on CGM hot gas from X-ray/UV absorption studies \citep{yao_dearth_2010}. Producing enough extended X-ray emission from purely thermal channels requires `tuning up' feedback to increase the thermal pressure at large radii, in turn breaking basic constraints on galaxy mass functions \citep{lau_x-raying_2025}.

Adding to this, recent observations suggest \textit{systematic suppression} of the thermal Sunyaev-Zel'dovich (tSZ) signal, or lower thermal energy, around $\sim$L$^{\ast}$ galaxies relative to the same simulations featuring thermal pressure-dominated, virialized halos \citep{das_thermal_2025}. If there is negligible non-thermal pressure, this presents a circum-galactic conundrum --- the extended X-ray surface brightness (XSB) profiles imply abundant \textit{hot} gas at large radii, with XSB $\propto \int \rm n_{e}^2\, T^{1/2}\, d\ell$, yet tSZ measurements probing the thermal pressure P$_{\rm th}$, with the relevant Compton-y parameter $y \propto \int \rm P_{\rm th} d\ell$ and total halo-integrated thermal energy $\tilde{Y}_{\rm 200} \propto  \int^{R_{\rm 200}}_{0}  P_{\rm th}\, dV$ indicate \textit{orders-of-magnitude} lower P$_{\rm th}$ than that of P$_{\rm th}$-supported halos from simulations and/or analytic theory. Reconciling this tension is paramount for any successful model of galaxy formation and feedback.

Most explorations of XSB profiles from simulated halos have focused on thermal emission via bremsstrahlung and metal line cooling. Recently, we proposed that the extended XSB profiles can very naturally be explained by inverse Compton up-scattering of CMB photons by CR electrons (hereafter, CR-IC; \citealt{hopkins_cosmic_2025}). Perhaps serendipitously, an $\mathcal{O}(1)$ fraction of $\sim$GeV CR leptons (which are plentiful in galaxies; \citealt{bell_cosmic_2013}) are expected to escape Milky-Way-like (MW-like) galaxies with fast effective transport speeds \citep{di_mauro_data-driven_2024} with their IC emission resulting in spatially extended $\sim$keV X-ray emission, with a spectrum mimicking that of a thermal continuum and the corresponding XSB profiles being consistent with the \citetalias{zhang_hot_2024} observations for empirical CR escape speeds.

Irrespective of feedback implementations, virtually all simulations explicitly modeling CRs with empirically-consistent transport parameters predict CR-pressure dominated halos \citep{salem_role_2016,butsky_role_2018,buck_effects_2020,Hopkins2020,rodriguez_montero_impact_2024}, which in turn modify CGM phase structure, as the CRs support \textit{cooler} gas, bringing models into better agreement with O/UV absorption measurements of ion column densities than non-CR counterparts \citep{Ji2020,butsky_impact_2022,thomas_why_2025,lu_constraining_2025}. Simultaneously, cooler CR-supported MW-mass halos naturally predict \textit{brighter} extended X-ray halos via CR-IC, which resemble \textit{hotter} gas (\citealt{hopkins_cosmic_2025}; hereafter \citetalias{hopkins_cosmic_2025}). As we discuss below, this reduces the tSZ signal, easing observational tensions.

In this Letter, we demonstrate using simple analytic arguments (\S\ref{sec:sec2}) the strong constraints on the CGM pressure budget provided by these recent observational and theoretical developments, \textit{heavily favoring CR-supported halos around $\sim L^{\ast}$ galaxies}. We then corroborate our analytic predictions with the latest mock observations of state-of-the-art, cosmological simulations of galaxy formation (\S\ref{sec:main_results}), and discuss our conclusions in broad context for galaxy formation (\S\ref{sec:discussion_conclusions}).

\section{Analytic Expectations for P$_{\rm th}$- or P$_{\rm CR}$-Supported Halos}\label{sec:sec2}

In this section, we lay out the basic qualitative scalings pertinent to the X-ray emission and tSZ properties of P$_{\rm th}$ and P$_{\rm CR}$-supported halos, respectively. We stress upfront that these are intentionally simplified, toy analytic models assuming steady-state conditions and spherical symmetry. Our aim is to establish these in order to juxtapose the qualitative difficulty or ease for each model class to meet the joint observational constraints. However, we will also note relevant results from state-of-the-art, hydrodynamical simulations of galaxy formation with and without explicit modeling of CRs to substantiate these simple arguments in full complexity.

\subsection{P$_{\rm th}$-Supported Halos}\label{sec:sec2.1}
For a virialized halo in thermal pressure equilibrium, hydrostatic balance is set by $\rho_{\rm gas} V_{c}^{2}/r = -dP/dr$, $P=n_{\rm gas}\,k_{B}\,T = \rho_{\rm gas} c_{s}^{2}$ where $P$ is the gas thermal pressure, $V_{c}$ is the circular velocity, $\rho_{\rm gas}$, $n_{\rm gas}$, and $T$ are the gas mass density, number density and temperature, $k_{B}$ is Boltzmann's constant, and $ c_{s}$ is the gas sound speed. To first approximation, we assume $\rho_{\rm gas} \propto r^{-2}$, a general scaling which is set by hydrostatic balance and is borne out by detailed simulations of galaxy formation in which $P_{\rm th}$ dominates (e.g., EAGLE, TNG, SIMBA, FIRE) with scatter generally manifesting around this scaling owing to varied feedback physics \citep[see][and references therein]{silich_x-ray_2025}. Here, we could choose a shallower $n_{\rm gas}$ profile owing to feedback which would increase the XSB at large R, though we stress this would not qualitatively change the tension of $P_{\rm th}$-dominated models with the tSZ data as it would change the relevant pre-factors but not the dimensional scalings with $n$ and $T$, as detailed further below.

The halo contains a baryon fraction $f_{b}$ within the virial radius R$_{\rm vir}$, which we define using the over-density criterion $\Delta_{\rm crit}(z=0) \sim 100$ \citep{bryan_statistical_1998} with the cosmic baryon fraction $\langle f_{b} \rangle = \Omega_b/\Omega_m$, where $\Omega_b ,\Omega_m$ are the dimensionless density parameters for baryonic matter and all matter, respectively. This results in n$_{\rm gas} = \frac{f_bM_{\rm vir}}{4\pi\mu m_p R_{\rm vir}} r^{-2}$. Hereafter, we specify explicitly when observables are evaluated at smaller radii (e.g., R$_{200}$). Now, assume $V_{c}(r)^{2} = V_{c,\,{\rm NFW}}^{2} (r\geq r_{\rm max}) + V_{\rm max,\,NFW}^{2}(r\leq r_{\rm max})$, i.e. the mass profile is a constant central $V_{c}$ within the galaxy, assigned to have $V_{c} = V_{\rm max}$ of the parent halo, and is an extended NFW mass profile with concentration $c_{\rm vir} \sim$ 10 beyond r$_{\rm max}$. 

Given these assumptions, we can solve for the temperature T$(r)$ and pressure profiles P$_{\rm gas}(r)$ and the corresponding total CGM-integrated thermal pressure following \citet{das_thermal_2025}, taking the same arbitrary angular diameter distance normalization of 500 Mpc:

\begin{equation}\label{eq:Y200_def}
        \tilde{Y}_{200} =  \frac{\sigma_{T}}{m_ec^2} E(z)^{-2/3} \int^{\rm R_{\rm 200}}_{0} P_{\rm gas}(r) 4 \pi r^2 dr/(500\, \rm{Mpc})^2
\end{equation}

which for our above assumptions gives 

\begin{align}\label{eq:Y200}
    \tilde{Y}_{200} \sim 1.25 \times 10^{-7} \rm{arcmin^2} \, E(z)^{-2/3} (\textit{f}_{b}/\langle \textit{f}_{b} \rangle) M_{\rm 200,12}^{5/3}
\end{align}

with M$_{\rm 200,12}$ the mass within R$_{\rm 200}$ in 10$^{12}$ M$_{\rm \odot}$ units.

Now, we assume that the maximum available metal supply is homogeneously mixed in the halo out to R$_{\rm vir}$. For this metallicity estimate, we take the peak star formation efficiency (M$_{\ast}$/f$_{\rm b}$*M$_{\rm halo}$) of $\sim 0.2$ around L$^{\ast}$, \citep{girelli_stellar--halo_2020} with up-to-date Type 1a SNe rates and the most optimistic yields \citep{maoz:Ia.rate,Leung2018}, which gives [Fe/H] $\approx -1$. Note that this permissive estimate does not account for metals which must be locked into stars via the known stellar mass-metallicity relation of galaxies. With this in hand, we can compute the expected total soft thermal X-ray (0.5-2 keV) luminosity, L$^{0.5-2 \,\rm keV}_{\rm X,\, CGM}$ = $\int^{R_{\rm 200}}_{0.5 R_{\rm200}} 4\pi r^2 \epsilon_X dr$, with the X-rays arising primarily from unresolved lines\footnote{$\epsilon_{X,\,{\rm line}}/({\rm erg\,s^{-1}\,kpc^{-3}}) \sim (Z/Z_{\odot}) \,10^{32}\,(n_{\rm gas}/10^{-5}\,{\rm cm^{-3}})^{2} \times f_{q}$, with $f_{q} = 4.4 \times q^{2}e^{-q}/(q+1)$ and an approximate scaling  of $q=0.6\,{\rm keV}/k_{B} T \approx 14.6/M_{200,\,12}^{2/3}$ for our assumptions here from fits to APEC \citep{mewe_calculated_1985,APEC_2001}.} for log$_{10}($M$_{\rm vir}) \lesssim$ 13.5 M$_{\odot}$, leading to  

\begin{align}\label{eq:LX_CGM_thermal}
    L^{0.5-2 \,\rm keV}_{\rm X,\, CGM}/10^{40} \rm{erg \, s^{-1}} \approx (\textit{f}_{b}/\langle \textit{f}_{b} \rangle)^{2} M_{200,\,12}\,\textit{f}_{q}\,\textit{f}_{profile}
\end{align}

where f$_{\rm profile}$ depends on the bounds of integration for ``CGM emission" and the emissivity profile --- f$_{\rm profile}$ = 1 integrated from 0 to R$_{\rm max, CGM}$, where R$_{\rm max, CGM}$ is often R$_{\rm 500/200/vir}$. Furthermore, the surface brightness from thermal emission S$_{\rm X,\, thermal}$ scales as S$_{\rm X} \propto$ r n$_{\rm gas}^2 f_q \propto r^{-3} f_q$, where $f_q$ introduces an exponential cutoff at r $\sim$ 0.5 R$_{\rm vir}$ due to declining T(r) as f$_q \propto e^{-0.5 \rm\, keV/k_BT}$, leading to an even steeper XSB profile at large $r$, in stark contrast to S$_{\rm X}$ $\propto r^{-1}$ profiles observed \citepalias{zhang_hot_2024}.

We note here that changing our assumption of the metallicity normalization would not qualitatively affect our conclusions on the joint tSZ \& X-ray constraints. For the volume filling, diffuse gas to have a higher metallicity than we chose would require a total baryon mass larger than that of f$_{b}$M$_{\rm vir}$, if we hold f$_{b}$ fixed. If we were to increase Z, this would require proportionally reducing f$_{b}$ in order to maintain agreement with the maximum possible mass of metals produced via SNe, which then reduces L$_{\rm X} \propto Z*n_{\rm gas}^2 * V$. Moreso than the already very optimistic assumption of the constant value of Z = 0.1 out to R$_{\rm vir}$ is the shape of the XSB profiles. To get S$_{\rm X}$ $\sim r^{-1}$, for any realistic CGM density profile which falls off as n$_{\rm hot\, gas} \propto r^{-\alpha_n}$ with $\alpha \in [1.5,3]$, S$_{\rm X} \propto r*Z*n_{\rm gas}^2$ means that Z(r) would need to steeply \textit{increase} with radius, which no model of metal propagation into the CGM or observations show. Virtually all simulations and UV absorption studies show negative radial gradients (i.e., Z(r) $\propto r^{-\alpha_{Z}}$, with $\alpha_{Z} \in [0.5-1.0]$) \citep[see][\S 3.2.1 and references therein]{hopkins_cosmic_2025}.

\subsection{P$_{\rm CR}$-Supported Halos}\label{sec:sec2.2}
\begin{figure}
    \centering
        \includegraphics[width=0.5\textwidth]{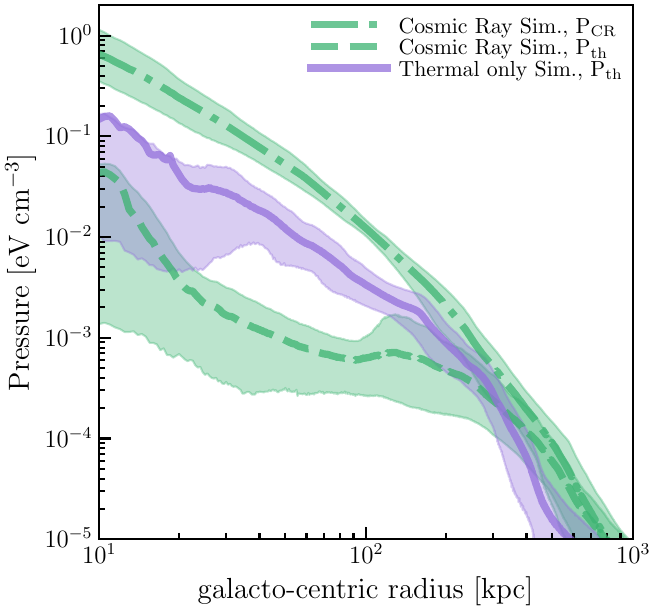}
        \caption{\textit{Radial profiles of P$_{\rm th}$ and P$_{\rm CR}$ in a simulated MW-mass halo.} P$_{\rm CR}$ (\textit{green dot-dashed}; \texttt{m12i} CR+) can replace P$_{\rm th}$ (\textit{green dashed}; \texttt{m12i} CR+) if CRs effectively escape the disk of the galaxy, and support the halo in contrast to halos supported primarily via P$_{\rm th}$ (\textit{purple solid}; \texttt{m12i} MHD+).}
        \label{fig:pressures}
\end{figure}

We now consider the alternative scenario of a CR-supported halo. Instead of the hydrostatic balance being set by the thermal pressure gradient, it is set by the CR pressure gradient $\nabla P_{\rm CR}$, which results in cooler gas ($T_{\rm gas} \sim \rm{few}\, \times  10^{4}\, K - 10^{5}\, K$ relative to the P$_{\rm th}$-dominated scenario at the same $\rho_{\rm gas} (r)$ for halo masses around M$_{\rm 200,12}$ \citep{Hopkins2020,Ji2020}. An example of the ensuing CGM thermal and CR pressure profiles for a simulated $M_{\rm 200,12}$ halo is shown in Figure \ref{fig:pressures}, though we refer the reader to \citet{Ji2020} for a relevant discussion of CGM phase structure. 

In these lower temperature halos, the emission/absorption properties are no longer set purely by collisional ionization equilibrium (CIE) temperature, but rather shift closer to the the photo-ionization equilibrium (PIE) temperature T$_{\rm PI, 5}$, which here we write in units of 10$^{5}$ K as an approximation to the range found by the CR-dominated simulations which self-consistently model thermal ionization in conjunction with photo-ionization from the UVB \citep{Hopkins2020,Ji2020}. 

In this scenario for the same assumptions for $n_{\rm gas}$ as in \S\ref{sec:sec2}, we find 
\begin{align}
    \tilde{Y}_{200} \sim 2.7\times10^{-8}\, \textrm{arcmin}^2 \,E(z)^{-2/3}(f_{b}/\langle f_{b} \rangle)\,M_{200,\,12}\,T_{\rm PI,\,5}
\end{align}
with the primary distinction from the P$_{\rm th}$-dominated scenario being a \textit{lower} $\tilde{Y}_{200}$ normalization (or total tSZ signal) by roughly an order of magnitude, and the power-law slope with M$_{200,\,12}$. 

Note for the fiducial model, we assumed a constant $T_{\rm PI,\,5} \sim 1 $, motivated by simulated CR-dominated halos at M$_{200,\,12}$ and $\rho_{\rm gas} \propto r^{-2}$, however, one could expect steeper power-law relations by instead reducing the halo density (or $f_{b}$) at fixed $T_{\rm vir}$, which is more likely for more massive halos with strong virial shocks wherein such rarefied gas would further cool ineffectively. 

In this case, the equilibrium scaling for $\rho_{\rm gas}(r)$ is instead set by the critical density condition of $\nabla P_{\rm CR} = \nabla \rho V_{\rm c}^2$, with $\nabla P_{\rm CR} \approx \frac{\dot{E}_{\rm CR}}{12\pi v_{\rm st, eff}r^2}$, where $v_{\rm st, eff}$ is the \textit{effective} CR streaming speed (containing streaming/advection-like terms, and assuming $\dot{E}_{\rm CR} \propto (\dot{M}_{\ast, \,R_{200}} + M_{\ast, \,R_{200}}) \propto f_bM_{\rm 200,12}^{\eta}$ where the subscript $R_{200}$ denotes contributions within R$_{\rm 200}$ (accounting for satellites), and $v_{\rm st, eff} \sim 60 \,\rm km\,s^{-1}$ is assumed. Note that v$_{\rm st, eff}$ is uncertain and determines the normalization of $\rho_{\rm gas}(r)$ in this approximation, so we take a constant, empirically motivated value at MW-mass here (but could in principle vary with radius as well as increase at higher mass scales) \citep{ruszkowski_global_2017,butsky_constraining_2023}. We also assume $\eta \approx 1$, which is roughly satisfied beyond the transition mass of M$_{\rm 200,12}$ once accounting for satellite contributions to $\dot{E}_{\rm CR}$ with increasing M$_{\rm 200,12}$. At M$_{\rm 200,12} < 1$, $\eta \approx 2$. This leads to a similar normalization as the fiducial model above, but retaining the self-similar scaling of $\tilde{Y}_{\rm 200} \propto M_{\rm 200,12}^{5/3}$:

\begin{align}\label{eq:y200_CR_selfsim}
    \tilde{Y}_{200} \sim 1.2\times10^{-8}\, \textrm{arcmin}^2 \,E(z)^{-2/3}(f_{b}/\langle f_{b} \rangle)\,M_{200,\,12}^{5/3}
\end{align}

For the equations above, we have scaled the behavior of CR-dominated MW-mass halos, but model predictions for the P$_{\rm CR}$ contribution in halos are strongly sensitive to uncertain CGM CR transport physics \citep{butsky_role_2018,hopkins_effects_2021,ruszkowski_cosmic_2023,lu_constraining_2025}. So these predictions can in a sense be considered \textit{rough lower-limits} to $\tilde{Y}_{\rm 200,CR}$ scalings, as $\tilde{Y}_{\rm 200,CR}$ would increase with lower non-thermal support from CRs.

We can now consider the X-rays arising from the expected CR-IC emission \citepalias{hopkins_cosmic_2025} from CRs injected via SNe (core collapse type II \& prompt Ia, as well as delayed Ia) and AGN-produced CRs streaming/diffusing into the CGM \citep{quataert_cosmic_2025,ponnada_hooks_2025,ponnada_2025_time_dependent}. We refer the reader to \citetalias{hopkins_cosmic_2025} for a more detailed presentation of the CR-IC formulation, but summarize the most relevant assumptions below for our analytic arguments.

To model CR injection, we start with the prompt Type Ia and core-collapse SNe rates of \citet{sukhbold:yields.and.explosion.props.dense.grid,hopkins_fire-3_2023}  and delayed Ia rates from \citet{maoz:Ia.rate}\footnote{$\dot{N}_{\rm type\, II\, \& \,prompt\, Ia} \sim 0.014\,(\dot{M}_{\ast}/M_{\odot})$, $\dot{N}_{\rm delayed\, Ia} \sim 1.5 \times 10^{-13}\,{\rm yr}^{-1}\,(M_{\ast}/M_{\odot})$, which are proportional to star formation rate (SFR; $\dot{M}_{\ast}$) and total stellar mass (M$_{\ast}$), respectively.}. At a given M$_{\ast}$, we invoke the SFR of the star-forming-main-sequence \citep{cooke:2023.sfr.main.sequence.evol}, and black hole accretion rate $\langle \dot{M}_{\rm BH} \rangle$ (averaged over many duty cycles owing to long CR transport timescales into the CGM for r $>>$ 10 kpc explored here) from \citet{torbaniuk:2024.bhar.sfr.mstar.relations}.

Each SNe is assumed to deposit 10$^{50}$ erg of total CR energy into the ISM, with $\sim$2\% of this total energy in leptons. We then assume a LISM-like spectrum of CRs escapes into the CGM (\citealt{bisschoff_new_2019}, already adjusted for leptonic losses within the dense magnetized disks/bulges; see \citealt[][for a detailed calculation]{ponnada_synchrotron_2024}) with approximately the same total lepton-to-total CRs ratio as, $f_\ell \sim f^\odot_\ell \sim$ 0.02 around the $\sim$GeV range of interest for soft X-ray CR-IC emission.

These escaping leptons then radiate away energy via IC losses as they stream/diffuse outwards, with $t_{\rm loss} \equiv p_{\rm cr}/\dot{p}_{\rm cr}^{\rm IC} \sim 1.2\,(1+z)^{-4}\,E_{\rm cr,\,GeV}^{-1}$\,Gyr and $t_{\rm travel} \equiv R/v_{\rm st,\,eff} \sim \,(R_{100})/(v_{100})$\,Gyr, where $v_{\rm st,\,eff}$ is the \textit{effective} streaming speed (which accounts for diffusive and streaming-like behaviors for a CR transport parameterization with constant $\kappa_{\|}$ and v$_{\rm st}$ common in the literature; \citealt[][]{ponnada_2025_time_dependent}), $v_{100} \equiv v_{\rm st,\,eff}(R) / 100\,{\rm km\,s^{-1}}$, and $R_{100}\equiv R/100\,{\rm kpc}$. 

Note that in the CR-IC model, before we introduce any of these simple assumptions for uncertain parameters (like $v_{\rm st,\,eff}$), to first order for \textit{any lepton spectrum peaked near $\sim$\rm{GeV},} $L_{\rm X,\, CGM, total}/10^{40} \rm{erg\, s^{-1}} \sim \int_0^{r_{loss}} e_{\rm CR,\ell}(r) 4\pi r^2 dr \sim \dot{E}_{40}$ out to the radius at which the $\sim$GeV leptons cool out of the $\sim$keV band and the XSB profile sharply truncates (where $\tau_{\rm loss} \gtrsim 2-3$).

Correspondingly, \citetalias{hopkins_cosmic_2025} provide a simple expression for the CR-IC produced XSB of $S^{\rm keV}_{X}/[\rm erg\, s^{-1}\, kpc^{-2}] \sim 10^{35.3} (1+z)^{4} (\dot{E}_{40})(R_{100})(v_{100}) \textit{f}_{\rm loss}$ 
where $\dot{E}_{40}\equiv\dot{E}_{\rm cr,\,\ell}/10^{40}\,{\rm erg\,s^{-1}}$, and
$\textit{f}_{\rm loss} = \rm exp[-\tau_{loss}]$, with $\tau_{\rm loss} \equiv t_{\rm travel}/t_{\rm loss} \sim E_{\rm cr,\,GeV}\,R_{100}\,(1+z)^{4}/v_{100}$. 

With the set of empirically-motivated assumptions above for the CR injection rate, composition, and transport, $L_{\rm X,\, CGM}$ is thus given by 
\begin{equation}\label{eq:LX_CGM}
\begin{split}
 L^{0.5-2 \,\rm keV}_{\rm X,\, CGM} / 10^{40}\,{\rm 
  erg\,s^{-1}} \sim 
  f_{\rm profile,\, CR-IC} [0.09 \dot{M}_{\ast}/({\rm M_{\odot}/yr})\\
  + 5700\,\epsilon_{-3}\,\langle \dot{M}_{\rm BH} \rangle/({\rm M_{\odot}/yr}) + 0.1\,M_{\ast}/(10^{11} {\rm M_{\odot}})] \\
\end{split}
\end{equation}
where $\epsilon_{-3}$ is the fraction of $\langle \dot{M}_{\rm BH} \rangle c^2$ injected into CR leptons, normalized to 10$^{-3}$. In comparisons below, we choose a $f_{\rm profile,\, CR-IC}$ corresponding to $v_{\rm 100} = 1,$ and $\dot{E}_{40}$ given by the bracketed term -- typically, $f_{\rm profile,\, CR-IC} \approx 0.5$. We stress that in contrast to the P$_{\rm th}$-dominated scenario, L$_{\rm X,\, CGM,\, CR-IC}$ is \textit{insensitive} to the density and/or temperature structure of the CGM, as it entirely depends on the CR lepton injection rate and transport into the CGM.
\vspace{-.5em}
\section{X-ray \& tSZ measurements together strongly favor CR-supported halos}\label{sec:main_results}

We now confront these analytic expectations with the latest observational tSZ and resolved X-ray emission measurements from stacks of $\gtrsim M_{\rm 200,12}$ halos.

\begin{figure*}

    \includegraphics[width=.8\textwidth]{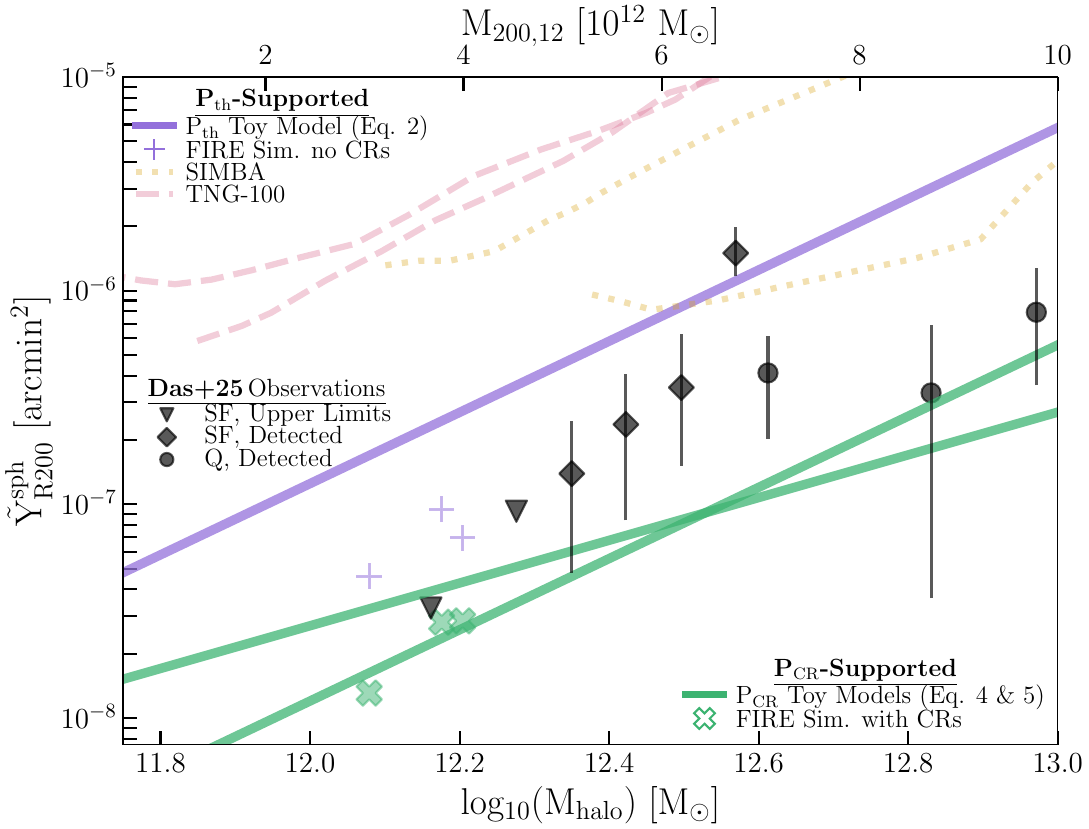}
        \centering
    \caption{\textit{Analytic and simulation predictions} for $\tilde{Y}_{\rm 200}$ vs. log$_{\rm 10}$(M$_{\rm halo}$) or M$_{\rm 200, 12}$. The analytic prediction for P$_{\rm th}$-dominated halos \textit{(purple solid)} lies factors of several above the upper-limits and star-forming or quenched detections at low and high M$_{\rm halo}$ of \citetalias{das_thermal_2025} (\textit{black triangles, diamonds \& circles}), whereas predictions for P$_{\rm CR}$-dominated halos \textit{(green solid)} can lie along the observations at low and high M$_{\rm halo}$, though the intermediate mass datapoints log$_{10}($M$_{\rm halo}) \in [12.3 - 12.6]$ are in between analytic predictions. Similarly, P$_{\rm th}$-dominated simulations like SIMBA and TNG \textit{(gold \& red, dotted and dashed)} generally over-predict $\tilde{Y}_{\rm 200}$ for quenched and star-forming  by \textit{over a dex}. Meanwhile, P$_{\rm CR}$-dominated FIRE simulations (\texttt{m12i,f,m} in order of M$_{\rm halo}$; \textit{green Xs)} are in agreement with upper-limits, in contrast to the same P$_{\rm th}$-dominated halos without CRs \textit{(purple pluses)}.}\label{fig:Y200}
\end{figure*}

In Figure \ref{fig:Y200}, we compare our analytic expectations for $\tilde{Y}_{\rm 200}$ vs. $M_{\rm 200}$ to the stacked detections from \citep{das_thermal_2025}. Immediately, two points are clear -- \textbf{first, halos which are primarily P$_{\rm th}$-supported \textit{systematically over-predict} the tSZ-inferred P$_{\rm th}$ measurements by $\gtrsim$ 1-1.5 dex.} This is emphasized by \citeauthor{das_thermal_2025} in their comparisons to large-volume simulations like TNG-100, SIMBA, and EAGLE \citep{pillepich_simulating_2018,dave_simba_2019,schaye_eagle_2015} at fixed $M_{\rm \star}$. Here, we convert the mean $M_{\rm \star}$ quoted by \citeauthor{das_thermal_2025} for the SIMBA and TNG galaxies using the relevant SMHM relations of \citet{cui_origin_2021} and \citet{pillepich_first_2018}, respectively. For the FIRE-2 simulations shown\footnote{\url{https://fire.northwestern.edu/}; FIRE-2  Halos: \texttt{m12i},   \texttt{m12f}, \& \texttt{m12m} CR+, MHD+ from the public DR2 \citep{FIRE_DR2}.}, we directly compute $\tilde{Y}_{\rm 200}$ within a projected radius of R$_{\rm 200}$, following Eq. \ref{eq:Y200_def}. 

We note that the P$_{\rm th}$-dominated MW-mass FIRE simulations lie close to the upper-limits on $\tilde{Y}_{\rm 200}$, in contrast to predictions from the large-volume simulations and thermal analytics. We attribute this primarily to FIRE lacking an AGN feedback model which would contribute to the overall halo pressure, and these halos primarily being in ``inflow modes" (rather than hydrostatic; see \citealt{hopkins_cosmic_Mpc_2021}) but as we show later in detail, this means they concomitantly under-predict X-rays.

Furthermore, the comparison between FIRE and the stacked observations qualitatively differs from the comparison to the large-volume simulations in \citet{das_thermal_2025} -- here we simply show a few publicly-available $\sim$MW-mass halos simulated with different physics to highlight the systematic difference in predictions for thermal- vs. CR-supported halos, but this means we are comparing against a very small sample of galaxies which neither represents the broad range of galaxies in the observations nor the effects of stacking. The comparisons to SIMBA and TNG, on the other hand, are averages of wide populations which feature a range of M$_{\rm halo}$ at fixed M$_{\rm \ast}$, which are more akin to the observations both in sample and stacking methodology.

\textbf{Secondly, halos which are P$_{\rm CR}$-supported \textit{can lie} surprisingly \textit{along} the measurements, in agreement with the upper-limits} at the lowest halo masses at $\sim M_{\rm 200,12}$.\footnote{With the exception of the datapoint near $M_{\rm 200} = 10^{12.5} \rm M_\odot$, where the inferred halo mass can be strongly sensitive to the invoked SMHM relation \citep{popesso_perils_2024}.} As we detailed earlier, many studies have shown SNe-injected CRs alone can result in P$_{\rm CR}$-dominated halos around $\sim 1-5\, \rm M_{\rm 200,12}$, and more recent developments demonstrate that even very small, fixed fractions of the AGN accretion energy being converted into CRs ($\epsilon_{\rm -3} \sim 0.1 - 10$), for reasonable CR transport parameters, can result in P$_{\rm CR}$-supported halos at $\gtrsim 5\, \rm M_{\rm 200,12}$ as well \citep{su_unravelling_2024,quataert_cosmic_2025,ponnada_2025_time_dependent}.

\begin{figure*}[htbp]
    \centering

    \hfill
    % Panel (b)
    \subfigure{%
        \includegraphics[width=0.49\textwidth]{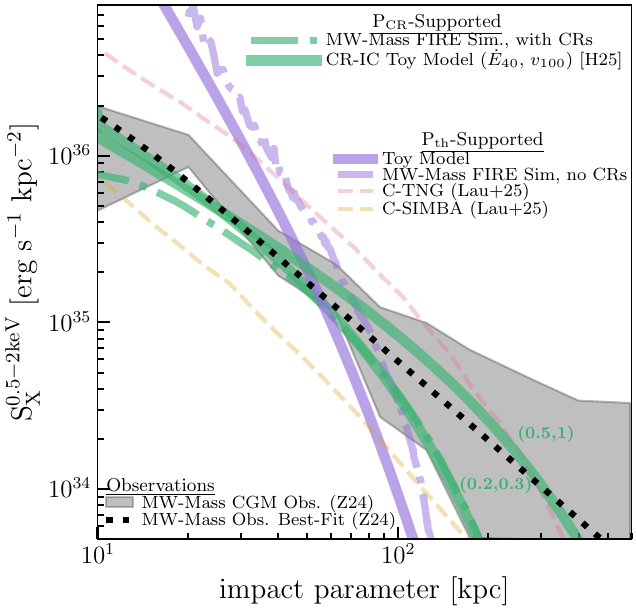}%
        \label{fig:XSB}
    }
    \hfill
    % Panel (c)
    \subfigure{%
        \includegraphics[width=0.49\textwidth]{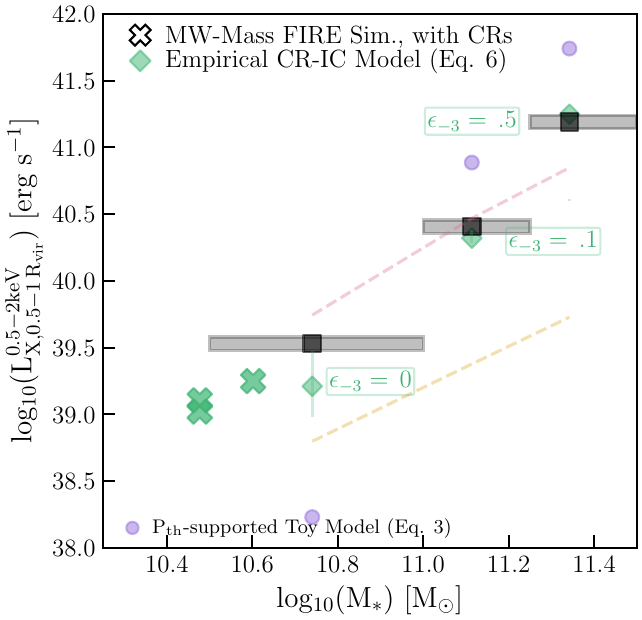}%
        \label{fig:LX_CGM}
    }
    \caption{\textbf{Left:} \textit{Analytic and Simulated XSB radial profiles in MW-mass halos.} The analytic XSB profile for a P$_{\rm th}$-supported halo (\textit{purple solid}) truncates steeply relative to the \citetalias{zhang_hot_2024} observations \textit{(black shaded and dotted)}, similarly to the P$_{\rm th}$-supported FIRE run \textit{(purple dot-dashed)}, whereas fiducial C-TNG \textit{(pink dashed)} can match the outer XSB shape and C-SIMBA \textit{(gold dashed)} vastly under-predicts the XSB. P$_{\rm CR}$-supported analytic predictions (\textit{green solid}, with variations in [$\dot{E}_{40},v_{100}$] labeled) can agree with the observed constraints, as also shown by a P$_{\rm CR}$-dominated simulation \textit{(green dot-dashed)}. \textbf{Right:} \textit{Total outer CGM soft L$_{\rm X}$, for $\sim L^{\ast}$ galaxies,} here integrated between 0.5-1 R$_{\rm vir}$. P$_{\rm CR}$-dominated MW-mass simulations \textit{(green Xs)} and empirical CR-IC models with varied AGN leptonic injection $\epsilon_{-3}$ \textit{(green diamonds, error-bars showing 0.3 dex scatter on $\dot{M_{\ast}}$)} agree with the observed \citetalias{zhang_hot_2024} best-fit \textit{(black squares, bands showing stacked range in $M_{\ast}$)}, while in agreement with $\tilde{Y}_{\rm 200}$. P$_{\rm th}$-supported analytic predictions \textit{(purple circles)} under-/over-predict L$_{\rm X}$ at low/high M$_{\rm \ast}$, C-TNG meets MW and M31-mass constraints but under-predicts at 2M31 mass, whereas C-SIMBA universally underpredicts L$_{\rm X}$. P$_{\rm th}$-supported models and simulations over-predict $\tilde{Y}_{\rm 200}$, irrespective of L$_{\rm X}$ predictions (c.f. Fig. \ref{fig:Y200})\textit{particularly at low/high masses}, but less so at intermediate M$_{\rm halo}$}.
    \label{fig:three_panels}
\end{figure*}

Turning our attention to the XSB profiles in Figure \ref{fig:XSB}, immediately two conclusions become clear: 1) the analytic expectations for the P$_{\rm th}$-dominated halos emission profiles are significantly truncated beyond R $\sim 100$ kpc, or 0.5 R$_{\rm 200}$ for M$_{\rm halo} \sim \rm M_{\rm 200, \, 12}$, and 2) the expected emission profile for a CR-IC profile matches \textit{surprisingly well} across the entire radial range probed by the observations. This is evinced in full generality by simulations -- comparing to a MW-mass simulation from the FIRE-2 simulation suite with and without a CR-dominated halo,shown in \citetalias{hopkins_cosmic_2025} as well. 

We also compare to the predictions from the fiducial CAMELS-TNG and CAMELS-SIMBA simulations \citep{villaescusa-navarro_camels_2021} as presented in \citet{lau_x-raying_2025}, hereafter referred to as as ``C-TNG" and ``C-SIMBA" respectively for brevity. Note, the fiducial C-SIMBA and C-TNG simulations evolve identical physical feedback prescriptions as their eponymous counterparts, but typically at lower resolution. 

As discussed and shown in \citet{lau_x-raying_2025}, the C-SIMBA XSB predictions systematically fall under the \citetalias{zhang_hot_2024} constraints, whereas the C-TNG XSB prediction can match the observations within 2$\sigma$. However, \citeauthor{lau_x-raying_2025} show that these two fiducial models \textit{fail} to match the observations at higher halo masses, requiring stronger feedback to reproduce the X-ray constraints, in turn unsuitably violating the M$_{\ast}$-M$_{\rm halo}$ relation. 

It is clear from comparing Fig. \ref{fig:Y200} and Fig. \ref{fig:XSB} that flattening the XSB profile at large radii from thermal emission necessarily means \textit{increasing} $\tilde{Y}_{\rm 200}$, bringing even the putatively agreeable TNG model at these M$_{200,12}$  further in disagreement with tSZ constraints relative to C-SIMBA. In short, this presents the primary contradiction for any self-consistent, thermal model for reproducing the observations -- reproducing the X-rays with stronger, traditional thermal and/or kinetic feedback pushing baryons out to larger scales ($\gtrsim 0.5 R_{\rm vir}$) in turn violates cosmological constraints while also boosting $\tilde{Y}_{\rm 200}$, whereas \textit{these contradictions can be resolved with non-thermal feedback and emission via cosmic rays}.

We exemplify this point further in Figure \ref{fig:LX_CGM}, where we specifically compare the total integrated L$_{\rm X}$ arising from the outer CGM ($r =  0.5-1\, \rm{R}_{\rm vir}$) to the same from the best-fit models of \citetalias{zhang_hot_2024}, focusing in on the radial range where this tension between models and observations largely appears to manifest. While other recent model comparisons to \citetalias{zhang_hot_2024,zhang_hot_2024-1} have focused on the halo emission out to smaller radii (R$_{\rm 500}$ or R$_{\rm 200}$; \citealt{lu_constraining_2025,lau_x-raying_2025,oppenheimer_introducing_2025}, we stress that for the very shallow XSB $\sim r^{-1}$ profiles observed by eROSITA\footnote{Note, naively propagating the parameter uncertainties on the spherical $\beta$-profiles in \citetalias{zhang_hot_2024}, the integrated emission in this outer radial range could \textit{maximally} vary by over an order-of-magnitude, particularly for the MW and M31 mass bins. But, this assumes the fit parameters are independent, and $\beta$-model parameters are notoriously degenerate \citep{mohr_properties_1999,kafer_toward_2019}. Moreover, such a maximal errorbar contradicts published values on the integrated L$_{\rm X}$ to R$_{\rm 500}$ in \citetalias{zhang_hot_2024-1}. So, lacking information for a proper error estimate, we neglect errorbars here. Ultimately, this does not affect our conclusions regarding the conundrum of meeting the joint tSZ \& X-ray constraints with purely thermal emission \& halo support.}, a significant fraction ($\gtrsim$ 50\%) of the integrated halo L$_{\rm X}$ within $\rm{R}_{\rm vir}$ arises from these outer radii, and it is particularly in cylindrical annuli beyond $\sim 100\, \textrm{kpc}$ that it becomes exceedingly difficult to match the observable constraints via thermal emission \citep[see][for a very detailed discussion of the fine-tuning problem for models meeting these X-ray constraints]{oppenheimer_introducing_2025}. In this plane, we show that the CR-IC model can explain the halo L$_{\rm X}$ \textit{while satisfying tSZ constraints}, with the share of CR-IC coming from AGN-produced CRs increasing at higher masses (M$_\ast \gtrsim 10^{\rm 11} M_\odot$), here chosen to `fit' the best-fit observational constraints for a CR-IC model with $v_{100} =1$ (still, with very small leptonic injection efficiencies relative to bright radio galaxies; \citealt{mcnamara_heating_2007}).

This lies in contrast to the C-SIMBA and C-TNG predictions, which we integrate using the `fiducial' XSB profiles of \citet{lau_x-raying_2025}, and our analytic predictions in \S \ref{sec:sec2}. These either systematically under-predict L$_{\rm X}$ from large radii in the case of C-SIMBA, or roughly match the \citetalias{zhang_hot_2024} constraints apart from the `2M31' stellar mass bin. Though \citet{lau_x-raying_2025} discuss in detail that the fiducial C-TNG model under-predicts in the L$_{\rm X}$ - M$_\ast$ plane when integrated solely within R$_{\rm 500c}$. Similar problems are found for our analytic predictions, where in the lowest mass bin the extended L$_{\rm X}$ is vastly under-predicted whereas at higher masses can be over-predicted, and for P$_{\rm th}$-dominated MW-mass FIRE simulations, which lie below the bounds of the plot. 

Remarkably, despite spanning nearly two orders-of-magnitude in L$_{\rm X}$ from large radii at each bin stellar/halo mass bin across $M_{\rm 200, \, 12} \in [1,10]$, these P$_{\rm th}$-dominated models \textit{ubiquitously} over-predict $\tilde{Y}_{\rm 200}$ relative to the \citet{das_thermal_2025} detections and upper-limits at MW-mass and the highest M$_{\rm 200, \, 12}$, though less so at intermediate masses as shown in Fig. \ref{fig:Y200}. This large spread in L$_{\rm X,\, CGM}$ predictions owes primarily to the strong sensitivity of the XSB cutoff radius to $T_{\rm gas}$ in the outer halo for \textit{any} model where L$_{\rm X}$ originates thermally. 

\begin{figure*}[ht!]
\centering
    \includegraphics[width=.75\textwidth]{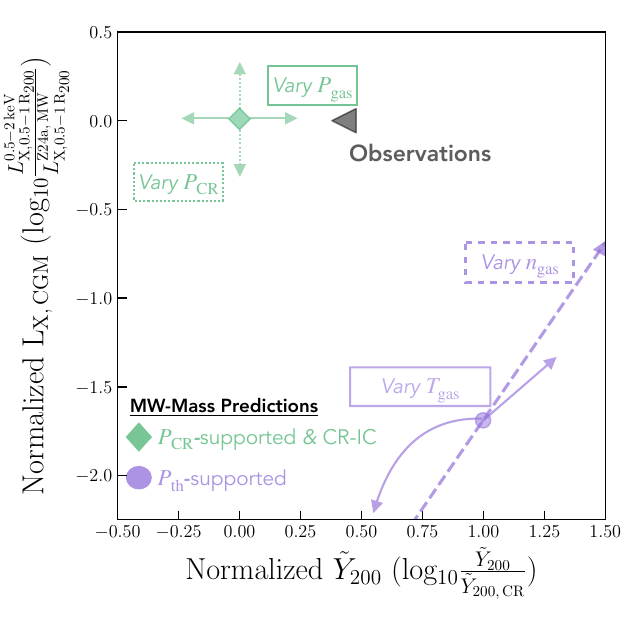}
        \centering
    \caption{\textit{Joint X-ray and tSZ observations constrain CGM Pressure: log$_{10}$ L$_{\rm X,\, CGM}$ vs. log$_{10}$ $\tilde{Y}_{200}$ for a MW-mass halo at $z=0.08$ (M$_{\rm 200,\,12} = 1.26$)}, here normalized to the integral of the best-fit MW-mass XSB profile of \citetalias{zhang_hot_2024} between 0.5-1 R$_{\rm 200}$ ($\approx$100-200 kpc) and to our $\tilde{Y}_{200}$ predictions for a P$_{\rm CR}$-dominated halo (Eq. \ref{eq:y200_CR_selfsim}), respectively. The L$_{\rm X}$ prediction for a P$_{\rm CR}$-dominated halo (\textit{green diamond}) with $\dot{E}_{40} = 0.66,\, v_{100} = 1$ lies close to the \citetalias{zhang_hot_2024} best-fit value, while in agreement with the tSZ upper-limit at M$_{\rm 200,\,12} = 1.25-2.59$ from \citet{das_thermal_2025} \textit{(black triangle)}, while the P$_{\rm th}$-dominated halo (\textit{purple circle}) is over 1 dex too high in $\tilde{Y}_{\rm 200}$. Varying $n_{\rm gas}$ or $T_{\rm gas}$ (\textit{purple dashed and solid lines}) necessarily move P$_{\rm th}$-dominated in the \textit{wrong directions} relative to the observational constraints, whereas changing P$_{\rm CR}$ or P$_{\rm gas}$ (\textit{green dotted and solid annotations}) move P$_{\rm CR}$-dominated model predictions in orthogonal directions in this plane, allowing for models to meet observational constraints.}\label{fig:tsz_xray}
\end{figure*}

We now illustrate the strong constraints placed by the tSZ \& X-ray observations on CGM properties in Figure \ref{fig:tsz_xray}, where we plot L$_{\rm X,\, CGM}$ integrated between 0.5-1 R$_{\rm 200}$ against the expected $\tilde{Y}_{200}$ for our analytic predictions for a M$_{200,12}$ (MW-mass) halo. Here, we normalize L$_{\rm X,\, CGM}$ to that from the best-fit MW-mass XSB profile of \citetalias{zhang_hot_2024} and $\tilde{Y}_{\rm 200}$ to the expectations of a CR-supported halo (Eq. \ref{eq:y200_CR_selfsim}). We show the upper-limit placed on $\tilde{Y}_{200}$ at the lowest M$_{200,12}$ bin in \citet{das_thermal_2025}, which corresponds roughly to the median M$_{200,12}$ for MW-mass galaxies in \citetalias{zhang_hot_2024}. Since we compare against L$_{\rm X,\, CGM}$ at slightly inner radii as compared to R$_{\rm vir}$, the expected L$_{\rm X,\, CGM}$ lies higher than shown in Figure \ref{fig:LX_CGM}, which in this annulus can agree well with the \citetalias{zhang_hot_2024} constraint. Still, the expected $\tilde{Y}_{200}$ is over a dex higher than the upper-limits of \citet{das_thermal_2025}, compared to the surprisingly good agreement of a reasonable CR-IC model with $\dot{E}_{40} = 0.66,\, v_{100} = 1$. 

The problem with thermal models, however, is not solely the normalization of L$_{\rm X,\, CGM}$ or $\tilde{Y}_{200}$, but the emergent behavior of changing relevant quantities: even if we were to ignore the non-linear dynamical effects of doing so, scaling n$_{\rm gas}$ moves a given P$_{\rm th}$-dominated point along a power-law relation with a slope of 2 since L$_{\rm X,\, CGM} \propto n_{\rm gas}^2$ and $\tilde{Y}_{200} \propto n_{\rm gas}$, which quickly over-/under-shoot the constraints in L$_{\rm X,\, CGM}$ while maintaining or \textit{increasing} the tension in $\tilde{Y}_{200}$, and changing $T_{\rm gas}$ would either move points \textit{exponentially lower} in L$_{\rm X,\, CGM}$, or \textit{in the wrong direction} in the plane along a power-law of 1/2 as L$_{\rm X,\, CGM} \propto \rm exp[-keV/k_BT_{\rm gas}]T_{\rm gas}^{1/2}$. 

In contrast, a P$_{\rm CR}$-supported model, in addition to its ease in predicting empirically-consistent normalization (by construction via choice of model parameters in the schematic here for emphasis), features an orthogonal basis in this plane, with model predictions moving vertically with P$_{\rm CR}$ or horizontally with P$_{\rm gas}$ to first order, highlighting a qualitatively simpler path to meet the joint observational constraints. Indeed, these observables together place a strong, direct constraint on model predictions for P$_{\rm CR}$, which remain highly uncertain in the CGM due to the complexities of CR transport \citep{hopkins_testing_2021,ruszkowski_cosmic_2023,hopkins_review_2025}. 

Furthermore, this means joint tSZ \& X-ray observations not only support a P$_{\rm CR}$-dominated halo paradigm, but provide an exciting avenue to constrain vastly varying model predictions for CR feedback, as opposed to synchrotron or $\gamma-\rm ray$ detections, which are degenerate with magnetic field strength or $\rho_{\rm gas}$, respectively \citep{ponnada_synchrotron_2024,sands_2025_gamma_ray}. 

For instance, re-arranging Eq. \ref{eq:Y200} and inline equations in \S\ref{sec:sec2.2} gives $ \frac{(500\, \rm Mpc)^2m_ec^2}{4\pi\sigma_T}\tilde{Y}_{\rm 200} = \int^{R_{200}}_0 P_{\rm gas} r^2 dr$ and $ \frac{f_{\rm CR,total} \nu_{\rm obs}}{12\pi}L_{\rm X, CR-IC} = \int^{R_{200}}_0 P_{\rm CR} r^2 dr$, where $f_{\rm CR,total}$ is $e_{\rm CR,total}/e_{\rm CR, \ell}$ and $\nu_{\rm obs} \approx \rm keV/\textit{h}$ . For SNe driven CRs with LISM-like spectra, $f_{\rm CR,total} \sim 20-100$,\footnote{Here, $f_{\rm CR,total} \sim $ 20 corresponds to the spectrum-integrated ratio from MeV-TeV energies, whereas $f_{\rm CR,total} \sim $ 50-100 are oft-quoted for the $\sim$GeV range \citep{cummings_galactic_2016,orlando_imprints_2018,bisschoff_new_2019}. } but this ratio \textit{increases} once the $\sim\rm GeV$ CR leptons begin to cool effectively to CR-IC in contrast to negligible $\sim\rm GeV$ hadronic losses. So L$_{\rm X, CR-IC}$ effectively sets a \textit{lower-limit} to the CGM-integrated CR energy and likewise S$_{\rm X, CR-IC}$ sets a lower-limit to e$_{\rm CR,tot}$. At masses where the lepton contribution from AGN might be more significant, this factor $f_{\rm CR,total}$ becomes more uncertain, but the lower-limit still remains robust. 

The strongest constraint from the present observations is thus at MW-mass. Taking the upper-limit on $\tilde{Y}_{\rm 200}$ at M$_{\rm 200,12} = 1.45$ from \citet{das_thermal_2025} and the interpreting the best-fit \citetalias{zhang_hot_2024} MW-mass L$_{\rm X}$ between 0.5-1 R$_{\rm 200}$ as CR-IC with a conservative estimate for $f_{\rm CR,total} = 20$ gives a lower limit of $E_{\rm CR, CGM-integrated} \gtrsim 1.2 \,E_{\rm th, CGM-integrated}$. Here, we assumed $\sim 50\%$ of the $\tilde{Y}_{\rm 200}$ signal arises from the same outer halo radii, which is approximately true for any NFW-like P$_{\rm gas}$ profile, but note we have not made any assumption regarding how the CRs reduce P$_{\rm gas}$ (via $n_{\rm gas}\, \rm{or} \, T_{\rm gas}$) relative to a P$_{\rm th}$-supported halo. In other words, the MW-mass observations already imply \textit{there is at least} as much CR energy in the CGM as in the thermal gas, or averaged over the outer halo volume, $\langle P_{\rm CR, CGM}\rangle \geq 1.2\, \langle P_{\rm gas, CGM} \rangle$. If new detections reveal $\tilde{Y}_{\rm 200}$ to be lower than the current upper-limits, this only furthers the indication of CR support.

\section{Discussion and Conclusions}\label{sec:discussion_conclusions}
In this work, we have synthesized findings from the latest X-ray \& tSZ measurements, simulations of galaxy formation \& corresponding mock observables, and analytic arguments, wherein detailed comparisons reveal a significant tension between model predictions and observations for the thermal state of halos around $\sim L^\ast$ galaxies. Taking the observations at face value, we find that they together already provide \textit{strict distinction between P$_{\rm th}$- and P$_{\rm CR}$-supported halos}. Several aforementioned works delineate the difficulty of matching the X-ray constraints \textit{alone} with traditional modes of stellar + AGN feedback \citep{lau_x-raying_2025,silich_x-ray_2025,oppenheimer_introducing_2025,zhang_tracing_2025}, a problem which only becomes more severe once the latest tSZ constraints are taken into account.

In brief, we stress again that it is quite fraught to meet the tSZ and X-ray constraints in tandem with a thermal explanation alone, as matching the X-ray constraints within R$_{\rm vir}$ requires boosting $f_{\rm b}$, n$_{\rm gas}$, or T$_{\rm gas}$ at outer radii, which in turn \textit{must increase $\tilde{Y}_{\rm 200}$}! To meet the X-ray constraints across the MW-2M31 mass range, matching both tSZ and X-ray requires reducing the halos' thermal pressure, in turn creating a fine-tuning problem for reproducing the X-rays at $M_{\rm 200, \, 12} \gtrsim 5$. This is particularly true for the MW-mass halos, where the discrepancy between models and observations for $\tilde{Y}_{\rm 200}$ is most severe. For instance, taking the P$_{\rm th}$-dominated FIRE runs (which lack AGN feedback, in contrast to C-SIMBA and C-TNG) and boosting the X-ray emission would shift those points upwards in $\tilde{Y}_{\rm 200}$ (Fig. \ref{fig:Y200}), exacerbating the tension. 

 We stress here that we do not make the case for CR-supported halos in this work via X-ray or tSZ constraints in isolation -- instead we highlight the powerful constraint they place on halos when considered in tandem. Below, we discuss relevant caveats and considerations of our conclusions, and highlight paths for future work which would expand upon the arguments made herein.

\subsection{Caveats}
\subsubsection{Stacking}
One immediate caveat is that the X-ray stacks of \citetalias{zhang_hot_2024} and the tSZ stacks of \citetalias{das_thermal_2025} are disparate samples, with the former extending only out to $z \sim 0.2$ in contrast to the latter's extending to $z \sim 1.2$. A matched sample for stacking analysis of tSZ and extended X-ray emission would be ideal, and we emphasize that we have demonstrated how such a sample could place extremely strong constraints on the physical state of the CGM around low-$z$ $\sim L^\ast$ galaxies.

However, we note here that since both observational samples are stacks, their biases would \textit{qualitatively} go in the same direction -- stacked, average quantities are sensitive to the brightest individual detections in a given bin, so to minimize sample bias, one ought to compare stacked values of $\tilde{Y}_{200}$ to stacked X-ray observations as done here. Though of course, tSZ maps are sensitive to n$_{\rm gas}$ whereas X-ray maps are sensitive to n$_{\rm gas}^{2}$ and suffer different noise biases. With this caveat and unless the clumping factors in the CGM $\langle n^2\rangle/\langle n \rangle$ happen to be quite high, the general trend for stacked values should be that stacked samples biased towards brighter X-ray sources should track brighter tSZ sources as well.

Moreover, as both \citetalias{zhang_hot_2024} and \citetalias{das_thermal_2025} note, these novel X-ray and tSZ measurements may suffer from large systematic uncertainties associated with difficult and distinct background and/or foreground subtraction. This is particularly true when attempting to infer gas temperature or density properties from either the X-ray and tSZ data -- and so we have focused on comparisons of \textit{observed} quantities herein. That being said, we caution the reader again regarding such large uncertainties which my evolve in the advent of up-and-coming observatories which will feature better sample coverage and sensitivity, as well as complementary probes of the CGM via kSZ and X-ray micro-calorimeter observations.

\subsubsection{Do X-rays from CR leptons imply P$_{\rm CR}$-dominated halos?}
We note here a potential way out of a necessarily P$_{\rm CR}$-dominated halo at $\sim$MW-mass could be realized if CR leptons are primarily AGN-produced. Take for instance the fiducial FIRE-2 model without CRs, where $\tilde{Y}_{\rm 200}$ is broadly consistent with the observed constraints of \citep{das_thermal_2025} as well as stacked observations of nearby spiral galaxies \citep{bregman_hot_2022} -- even a small $\sim \mathcal{O}(10^{-5}-10^{-4})$ fraction of the expected $\langle \dot{M}_{\rm BH}\rangle$ at those masses \citep{torbaniuk:2024.bhar.sfr.mstar.relations} being converted into CR leptons could reproduce the detected X-rays via CR-IC. Though we caution that this possibility requires assuming the SNe-produced leptons contribute negligibly to the emergent CR-IC signal, which for our empirical assumptions above, we have demonstrated can entirely account for the X-rays at MW-mass. 

Nonetheless, this is an interesting scenario to consider, particularly at increasing halo mass ($\gtrsim$ M31-mass), where AGN contribution to the lepton budget appears to be required of the CR-IC model to reproduce the extended X-ray constraints instead of SNe-driven CRs alone. Though, in that regime, the X-ray constraints would still have to be reconciled with remarkably low $\tilde{Y}_{\rm 200}$ detections, again indicating P$_{\rm CR}$-support, but perhaps with a larger share arising from leptons vs. largely hadronic P$_{\rm CR}$ from SNe-driven CRs.

\subsubsection{What about known detections of hot halo gas?}
The extended X-ray emission originating from CR-IC we argue for herein \textit{does not preclude} known sources of hot thermal emission/absorption from UV/X-ray observations of the MW halo \citep[e.g.][]{yao_dearth_2010,kaaret_disk-dominated_2020,bluem_widespread_2022,gupta_thermal_2023,ponti_abundance_2023} -- we have purposefully focused our comparisons to the X-ray observations here on \textit{outer halo} radii where it is particularly difficult to explain extended X-ray emission with thermal models. 

By no means do we assert that observed, clearly thermal line emission/absorption from the inner halo must be CR-IC, and indeed in simulated CR-dominated halos, these features  thermal gas at smaller radii do appear  \citep{lu_constraining_2025}. Rather, we are arguing here that the soft band-integrated $L_{\rm X, CGM}$ emerging from large radii in $\sim$L$^\ast$ halos is likely non-thermal in nature. At higher halo masses (e.g. more massive groups and clusters) the thermal emission certainly bears increasing importance at large radii, though CR-IC may contribute and explain interesting phenomena within ``cool-cores'' \citep{hopkins_coolcores_2025}.

\subsection{Connecting Constraints and Future Work}
Beyond tSZ and X-rays, independent budgeting of cosmic baryons from the dispersion measure-redshift relation out to $z \sim 1$ \citep{connor_gas-rich_2025} indicates evidence for `baryon-evacuated' halos, with roughly $\sim9\%$ of cosmic baryons residing in virialized halos. As noted therein and by other works (e.g., \citealt{oppenheimer_simulating_2021}), this is similar to predictions from SIMBA and TNG, but as we detailed in this work, still remain at odds with the joint tSZ \& X-ray constraints.

Contrarily, CRs, which are required to escape galactic disks into halos by empirical constraints in the LISM \citep{strong_global_2010,di_mauro_data-driven_2024}, observations \citep{lacki_physics_2010}, and mock observational comparisons of simulations \citet{chan_cosmic_2019,hopkins_testing_2021,ponnada_synchrotron_2024,ponnada_synchrotron_2024-1,ponnada_hooks_2025,martin-alvarez_extragalactic_2024} naturally explain lower P$_{\rm th}$ simultaneously with extended soft X-ray emission, and push baryons out to larger scales \citep{hopkins_cosmic_Mpc_2021}. And indeed, the redshifts probed by these observations are \textit{precisely} where CRs are expected to be of significance in $L^{\ast}$ halos \citep{Hopkins2020}. Moreover, as we have shown, it is much simpler to conceive of a model which reconciles the joint constraints within a CR framework due to the qualitatively different dependencies of the observables on physical parameters. 

As we describe in \citetalias{hopkins_cosmic_2025}, the details of the XSB profiles when interpreted as CR-IC provide unique, unprecedented constraints on the effective, bulk CR transport speeds in galaxy halos. Past inferences of non-thermal pressure support in the CGM around MW-mass galaxies have been comparatively indirect, via modeling of absorption line ratios and column densities \citep{werk_cos-halos_2014,butsky_constraining_2023} -- in contrast, these novel tSZ and X-ray observations synchronously provide \textit{direct} evidence for CR-pressure support. Spatial cross-correlation of the Compton-y parameter in conjunction with XSB profiles may further provide powerful insights.

Additional joint tSZ \& X-ray observations may also provide distinctive constraints on the multi-channel energetics of AGN feedback and quenching, which an increasing number of studies suggest may be CR-related at the group mass scales of $M_{\rm 200, \, 12} \gtrsim 5$ \citep{wellons_exploring_2023,su_unravelling_2024}. Curiously, \citet{zhang_hot_2025} find that quiescent galaxies in the MW-2M31 mass range are \textit{brighter} in CGM X-ray emission than their star-forming counterparts, while \citet{das_thermal_2025} find low $\tilde{Y}_{\rm 200}$ for quiescent galaxies in this mass range. This hints at the possibility of CRs from AGN playing a role in quenching and modulating CGM thermal pressure and emission properties. More detailed constraints may help reveal \textit{how} AGN quench massive galaxies and maintain quiescence at low$-z$, and constrain models, as simulations featuring CRs from AGN with varied injection efficiencies predict vastly differing CGM properties (Goyal et al. in prep.).

Probing quenching and ensuing observable galaxy population properties, however, requires large volume cosmological simulations, as opposed to zoom-in or idealized simulations most common in the literature exploring CR effects on galaxy formation \citep{hopkins_review_2025}. Indeed, this front is progressing via implementations of sub-grid models for CR feedback in big cosmological boxes \citep{ramesh_illustristng_2025}, and new theoretical advances capturing time-dependent effects which may be of non-linear importance for how CRs rearrange baryons and regulate galaxies \citep{ponnada_2025_time_dependent}. Such simulations will allow for cohesive and self-consistent observational comparisons of statistical galaxy populations to not only tSZ and X-ray observations, but also weak-lensing constraints \citep{des_collaboration_dark_2022,siegel_joint_2025} and independent constraints on the matter distribution from FRBs \citep{connor_gas-rich_2025,sharma_hydrodynamical_2025,sharma_probing_2025}. \newline

We thank the anonymous referee for helpful and constructive comments. SP thanks Eliot Quataert for helpful comments which improved this manuscript. Support for SP and PFH was provided by a Simons Investigator Award.
\bibliography{cgm_xray}{}
\bibliographystyle{aasjournalv7}



\end{document}